\pdfoutput=1
\documentclass[11pt]{article}
\usepackage[margin=1in]{geometry}
\usepackage{xcolor}
\usepackage[hidelinks]{hyperref}
\usepackage{graphicx}
\usepackage{makecell}
\usepackage{tabularx,array}
\newcolumntype{Y}{>{\centering\arraybackslash}X}

\title{Designing a Secure and Resilient Distributed Smartphone Participant Data Collection System%
\thanks{This work was supported by the National Cancer Institute of the National Institutes of Health under award number R37CA276365. The content is solely the responsibility of the authors and does not necessarily represent the official views of the National Institutes of Health.}}

\author{
Foad Namjoo$^{1}$ \and
Neng Wan$^{2}$ \and
Devan Mallory$^{3}$ \and
Yuyi Chang$^{3}$ \and
Nithin Sugavanam$^{3}$ \and
Long Yin Lee$^{2}$ \and
Ning Xiong$^{2}$ \and
Emre Ertin$^{3}$ \and
Jeff M. Phillips$^{1}$
}

\date{}

\begin{document}
\maketitle

\begin{center}
\small
$^{1}$ Kahlert School of Computing, University of Utah\\
\texttt{foad.namjoo@utah.edu, jeffp@cs.utah.edu}\\[4pt]
$^{2}$ School of Environment, Society \& Sustainability, University of Utah\\
\texttt{neng.wan@utah.edu, u1467261@utah.edu, ning.xiong@utah.edu}\\[4pt]
$^{3}$ Department of Electrical and Computer Engineering, The Ohio State University\\
\texttt{mallory.115@osu.edu, chang.1560@osu.edu, sugavanam.3@osu.edu, ertin.1@osu.edu}
\end{center}

\begin{abstract}
Real-world health studies require continuous and secure data collection from mobile and wearable devices. We introduce \textbf{MotionPI}, a smartphone-based system designed to collect behavioral and health data through sensors and surveys with minimal interaction from participants. The system integrates passive data collection (such as GPS and wristband motion data) with Ecological Momentary Assessment (EMA) surveys, which can be triggered randomly or based on physical activity. MotionPI is designed to work under real-life constraints, including limited battery life, weak or intermittent cellular connection, and minimal user supervision. It stores data both locally and on a secure cloud server,
with encrypted transmission and storage.
It integrates through Bluetooth Low Energy (BLE) into wristband devices that store raw data and communicate motion summaries and trigger events.   
MotionPI demonstrates a practical solution for secure and scalable mobile data collection in cyber-physical health studies.

\end{abstract}

\noindent\textbf{Keywords—} System Design; Cyber-Physical Systems; Mobile Health; Wearable Sensors

\section{Introduction}
\label{sec:intro}

Understanding and improving personal health behavior is a major challenge.  Although advances in smart devices make it possible to capture fine-grained medical, personal, and movement data about individuals, doing so as part of a medical study still presents numerous challenges.  The resulting system must be inexpensive, automated, easy-to-use, reliable, and protect private patient data.  This requires a carefully designed secure smartphone system.  

General personalized health studies require multiple components.  First, \emph{movement} is a key indicator of fitness and activity, and its relationship to health outcomes is an important factor.  As such, smart systems need to collect various high-frequency statistics, such as GPS location and tracking of wrist movement and heart rate through peripheral smart devices.  
Beyond just tracking, many health studies require \emph{interventions}, where the smart app detects certain activities and prompts the study patient to perform a task.  This task may also collect data like a survey.  

To achieve these goals, the smartphone interface should be minimal and seamless to ensure that it is used and that participants only use the intended features.  It should be implemented efficiently with only the necessary Bluetooth connections to peripherals and other storage costs, so the batteries of the equipment can last a full day without recharging.

\begin{figure}[b!!]
    \includegraphics[width=\linewidth]{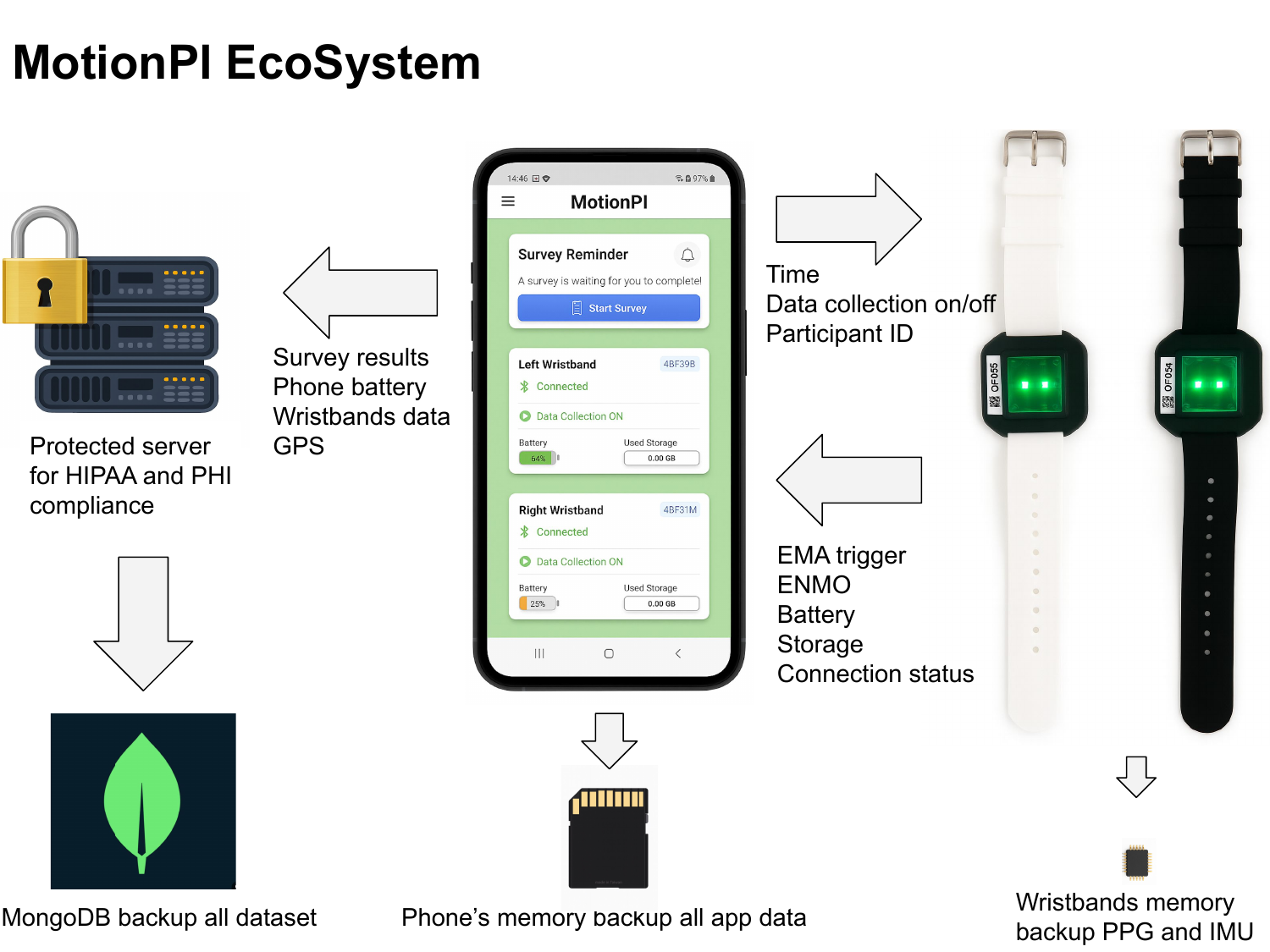}
    \caption{Diagram of the MotionPI Ecosystem.  The App plays a central role in coordinating the data collection from peripherals, high-frequency sensing, and EMAs, and its secure and distributed storage.  }
    \label{fig:diagram}
\end{figure}

When dealing with private patient health information, it must be secure.  The equipment provided to study participants should not have residual data kept on the device of previous participants and if it is there, it should be encrypted.  
The storage method should also be resilient and verifiably consistent.  These studies need many participants, and it is a complex process to recruit them and convince them to actively participate in the procedures.  If data is lost, it is a waste of time and expense.

\paragraph{Our solution.}
We provide a smartphone system that stores data in multiple ways, both in the smartphone storage (with encryption) and in a secure cloud environment.  
This setup enables distributed, simultaneous, and secure uploads directly into a NoSQL (MongoDB) database. This structure, shown in Figure \ref{fig:diagram}, generates many advantages:
(1) It is easy to monitor the data collection process and quality, 
(2) It enforces access control so that only authorized persons and all data analysts part of the project can access the data at any time, and
(3) It enables verification that all the collected data has been successfully uploaded by comparing them with the locally stored files. 
To accomplish this, our proposed system, MotionPI, needs to make several critical design decisions.  These include how the data are stored (both format and server environment), how to implement a secure container deployment, how to ensure data is eventually sent to the cloud when the smartphone's connectivity is sporadic, and it must ensure efficiency so it will not drain the smartphone battery too quickly.  
We document how we solve these issues in this report.  
Our system is currently being actively deployed in a large NIH-funded study.




\section{Related Work}
\label{sec:related}

Several mobile health data collection platforms have been developed to support behavioral and physiological studies in the real world. Among the most widely used are \textbf{mCerebrum}~\cite{hossain2017mcerebrum}, \textbf{Beiwe}~\cite{onnela2021beiwe}, and \textbf{RADAR-base}~\cite{ranjan2019radar}. Although mCerebrum is a widely used mobile health platform that supports sensor data collection and digital biomarkers, we encountered compatibility issues when trying to update and run the software on newer versions of Android. 

The \textit{mCerebrum} framework offers extensive sensing capabilities and supports high-frequency data collection, but its configuration requires technical expertise and often relies on continuous network connectivity. \textit{Beiwe} provides strong privacy guarantees by encrypting data before upload, but is designed primarily for centralized studies and lacks support for edge-level computation or offline-triggered surveys. \textit{RADAR-base} uses wearable devices and smartphones to stream data in real time, but its architecture assumes stable connectivity.

Other platforms, such as \textit{AWARE }~\cite{ferreira2015aware}, allow sensor-based logging, but offer limited support for secure local storage or adaptive EMA logic.

In contrast to these systems, our proposed platform, \textbf{MotionPI}, is specifically designed for deployments with limited network availability, minimal participant interaction, and high privacy requirements. MotionPI enables encrypted local storage, robust background uploading to a secure cloud server, and EMA scheduling logic triggered by both time and sensor thresholds.

MotionPI combines a \emph{dual storage} architecture (encrypted local and cloud redundancy) with \emph{timed EMA triggers}—including BLE-based physical-activity detection (based on ENMO~\cite{bakrania2016enmo})—and device-scoped authentication for offline operation that tolerates intermittent connectivity; see Table~\ref{tab:platform-compare}.  
The remainder is necessary engineering integration (phone app + wristbands + backend + ops) that makes the system usable in real deployments while preserving confidentiality and availability under weak networks.

\begin{table}[ht]
\centering
\caption{Comparison of mobile data collection platforms. MotionPI supports offline-triggered surveys, BLE-based activity detection, and dual storage (local + cloud).}
\label{tab:platform-compare}
\small
\setlength{\tabcolsep}{-2.5pt}
\renewcommand\arraystretch{1.27}
\begin{tabularx}{\linewidth}{lYYYY}
\hline
\textbf{Platform} &
\makecell[c]{\textbf{Timed}\\[-3pt]\textbf{EMA trigger}} &
\makecell[c]{\textbf{BLE}\\[-3pt]\textbf{activity trigger}} &
\makecell[c]{\textbf{Dual}\\[-3pt]\textbf{storage}} &
\makecell[c]{\textbf{Offline}\\[-3pt]\textbf{sign-in}} \\
\hline
mCerebrum~\cite{hossain2017mcerebrum} & yes & yes & no & no \\ Beiwe~\cite{onnela2021beiwe} & yes & no & yes & no \\
RADAR-base~\cite{ranjan2019radar} & yes & no & no & no \\ AWARE~\cite{ferreira2015aware} & yes & no & yes & via config URL \\ 
\textbf{MotionPI} & \textbf{yes} & \textbf{yes} & \textbf{yes} & \textbf{yes} \\
\hline
\end{tabularx}
\end{table}

\section{The Smartphone App}
\label{sec:app}

The MotionPI mobile application is a secure, resilient, and highly modular Flutter-based system developed to collect behavioral and physiological data in real-world health studies. It integrates passive (e.g., GPS and motion) and active assessments (surveys) through sensors and Bluetooth-connected wristbands.  The app operates continuously with minimal user interaction and handles limited connectivity and low-power environments. 
It automatically operates in the background to handle continuous operation and geographically localizes each timestamp. Each participant is assigned an ID used as a unique identifier for their data and masking their identity;  a study coordinator can maintain a list that maps these abstract IDs to participant names and information.  Similarly, the data are marked with a phone ID using the Android advertising ID. The app then authenticates with the secure cloud server using JSON Web Tokens (JWT). If a token is missing, the app fetches a new one by signing up.


The user can select the name of the wristband on the settings page, and then the app connects simultaneously to two MotionSense~\cite{kwon2021validity} wristbands (left and right). It automatically discovers, connects, and resubscribes to wristbands using stored Media Access Control (MAC) addresses, and it monitors battery, charging status, storage usage, and motion using Euclidean Norm Minus One (ENMO) values~\cite{bakrania2016enmo}. The app displays real-time indicators for wristband connection status, battery status, and used storage on the home screen. It stores all data locally and sends them securely to the cloud server.

\begin{figure} [ht]
    \centering
    \includegraphics[width=\linewidth]{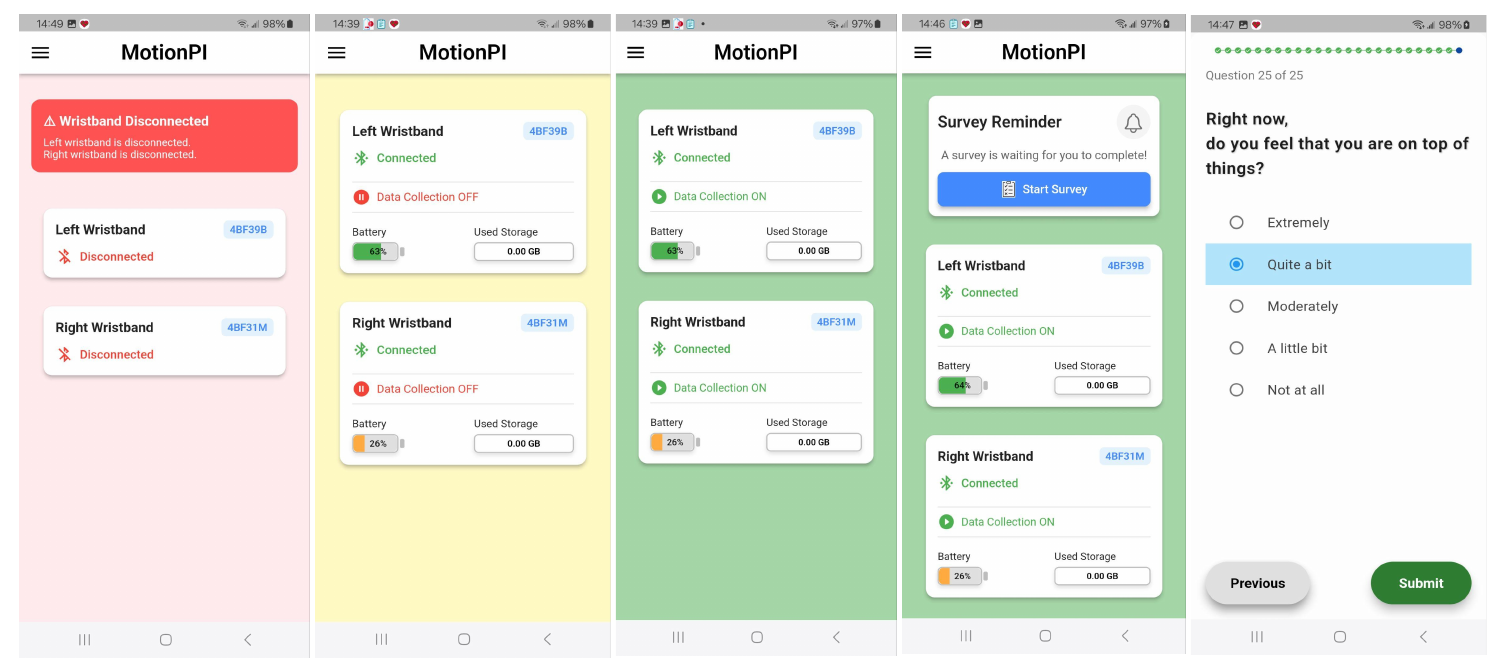}
    \caption{Screenshots from the MotionPI App.  The first (red) shows before the wristbands are connected, then after they are connected (yellow) but not activated, and in the middle, all are activated (green).  The right two show when an EMA survey is triggered and available, and then within the EMA survey.}
    \label{fig:screenshots}
\end{figure}

The app schedules data collection by enabling from 7:30 AM to 9:30 PM daily. During this time, green LEDs on the wristbands and a status message on the home screen (the green screens in Figure \ref{fig:screenshots}) confirm enacted data collection to the participant. 

GPS coordinates are encrypted with AES-256 (CBC mode) before storage, as this information can reveal the identity of the participant (e.g., by inferring their home and work address). 

The app delivers two types of EMA surveys via notification in a specific situation:
Random Surveys; it is triggered three times daily at random times within morning, afternoon, and evening blocks. Activity Surveys; it is triggered upon detection of physical activity via the ENMO threshold of the wristband; see Figure \ref{fig:screenshots}.  Participants will be directed step by step to submit the survey. The submitted survey data is saved locally and securely transmitted to the cloud database.

All major app actions are logged and sent to the MongoDB backend and are easy to query in MongoDB Compass for deeper analysis.
Not only does the system log the core data collection tasks, but also other critical events in the operation, such as when the participant turns Bluetooth on or off, when a wristband connects or disconnects from the app, and app state events, such as when data collection is enabled or disabled, when a wristband battery drops below 20\%.  Importantly, it records a variety of EMA survey-related events such as when an activity-triggered survey is initiated by wristband motion or otherwise, when an activity survey notification appears, when a participant declines the survey, and when a survey expires after 30 minutes without completion.  These are important to assess how well a participant is following the study requirements and to calculate their monetary compensation for doing so.  
All of the above logs include participantID, phone ID, timestamp, wristband MAC, and local time.

\subsection{Access Patterns for User Interface and Interaction}  
The app has different functionality for administrators (those running the study) and participants (the patients who are being monitored in the study).  

The MotionPI app user interface is designed for clarity, accessibility and minimal burden for all types of users, while offering essential features for both participants and administrators. The homepage presents real-time information, including wristband connection status, battery and storage levels, data collection state, and survey reminders; again, see Figure \ref{fig:screenshots}. If either wristband becomes disconnected, red banners and a short vibration appear to alert the participant immediately. The app will automatically connect whenever the wristbands return to connection distance or recharge.

The password-protected profile page allows users to set the participant ID (e.g., motionpi088), choose their dominant hand (left or right), and manually enable or disable data collection. These controls are critical for initial setup and troubleshooting.

Within the Settings page (also password protected), the admin can assign wristband labels with MAC addresses for wristband identification, update the app, and reset wristbands' internal memory. 
These settings are persistent, and allow for simple updates within large deployments.  


\section{Wristbands}
\label{sec:wristbands}


The study uses 
wristband sensors extending MotionSenseHRV~\cite{kwon2021validity} 
to facilitate tracking and storing the levels of physical activity of the participants and provide triggers for the MotionPI application. The newly-designed wristbands support Bluetooth Low Energy (BLE) communication (Nordic Semiconductor nRF5340) with 4GB on-board NAND flash (Micron Technology 4x MT29F8G01ADBFD12) for up to 30-day data recording capability with the sampling rate of 32 Hz and 64 Hz for IMU and PPG, respectively.  

The onboard sensors consist of a 9-axis Inertial Measurement Unit (IMU) (TDK InvenSense ICM20948) and a photoplethysmography sensor (PPG) (Analog devices MAX86141) that measures blood flow using green and infrared wavelengths (OSRAM Inc. SFH701). The sensor enclosure is manufactured using the 3D printing technique (Stratasys, Ltd.), taking a dimension of $42\times44\times 12$ mm, which is comparable to many commercially available wristbands. Additionally, the enclosure consists of optical barriers to mitigate ambient noise caused by variations in lighting conditions, improving the signal-to-noise ratio of the PPG signals. Finally, the intensity of the illumination is adapted to obtain PPG measurements that are agnostic to the participant-specific variations.

The device communicates with the MotionPI mobile application via a set of Bluetooth characteristics, which include commands to start or stop data collection, encode the participant ID, data collection time, and, when necessary, erase the onboard storage at the beginning of the study.  To reduce communication overhead, only summary statistics of the activity state are transmitted via Bluetooth notifications. This notification includes the flag that indicates the detection of moderate to vigorous physical activity (MVPA) over an epoch of the past $7$ minutes from the current time.

\subsection{Physical Activity Triggers}
The physical activity level is measured using accelerometry measurements. The IMU sensor measures wrist acceleration along with the effect of acceleration due to gravity. The IMU sensor also has a significant measurement bias. These imperfections in the sensor are addressed by constantly correcting for the bias error and compensating for acceleration due to gravity by subtracting the gravity vector. This quantity is referred to as Euclidean norm minus one (ENMO)~\cite{bakrania2016enmo}, which is an approximation of the linear acceleration. ENMO is averaged over intervals of $15$ seconds to obtain a measure of energy expenditure. We define a 15-second bout as MVPA if ENMO exceeds 0.1006~g (100.6~mG).
 We keep track of the MVPA levels over a $7$-minute epoch and declare the epoch as MVPA if there are $4.9$ minutes ($294$ seconds) in the last 7 minutes of bouts that were declared as MVPA. The wristband notifies the MotionPI application if an epoch is detected as MVPA. 

\subsection{File-System Implementation}

The sampled data, comprised of IMU and PPG measurements, is stored locally on the wristband using a FAT16 file system to organize the data. To improve performance and reliability, the file system was fitted with a custom-developed Flash Translation Layer (FTL), bridging the raw binary data into a mountable file system. Instead of storing all parts of the file system in a singular set of addresses like a conventional FTL, the custom FTL separates the file allocation table section and the data section on separate flash devices. Specifically, the file allocation table is stored on an onboard 8 MB NOR flash, while the data section is stored on the main 4 GB NAND flash. The file system was also modified to only support pre-allocated files, with timestamps and sizes of the file being marked only on file creation. This design for hardware separation of the file allocation table has several advantages - namely, it completely removes the need for file creation wear leveling on the NAND components, making it more durable (as those algorithms no longer need to run, thus eliminating the total number of write erase cycles needed for any file creation). While the system is slower during initial file creation (due to NOR being slower than NAND), writes to files themselves do not modify the file allocation table (due to the pre-allocation scheme), and due to this, the file system is faster than a traditional FTL when writing data for already opened files. Performance metrics and source code files for this FTL are available at \cite{msense2025}.

Most importantly, this method offers data protection benefits, since in order for the data to be grabbed, attackers will need to have prerequisite knowledge of the physical layer file system separation, and extract all data from both the external NOR and NAND. Once this is done, the attacker would also need to figure out the file allocation table location and virtual addressing scheme on the NOR, which is non-trivial due to the idea that the file allocation table is mixed in with other miscellaneous runtime data on the external flash. The storage is made accessible by USB and can be viewed after collection via an administrator's computer. From a high level, the data is stored contiguously in 4 KB files, with a single file being created for both IMU and PPG, respectively. A graphical tool 
\cite{yams2025}   
is developed to download and extract onboard data to a human-readable format.



\section{Secure, Robust Interaction with Cloud}
\label{sec:cloud}


The MotionPI mobile app communicates with a secure, resilient, cloud-hosted backend. 
The backend is implemented using the Express.js framework for the MotionPI data storage, authentication, and API management architecture. The cloud infrastructure ensures persistent, highly available storage and facilitates centralized data analysis, monitoring, and system scalability. 

This secondary cloud storage is implemented using a Podman-managed MongoDB container that is running on a Rocky Linux virtual machine. 
The container is configured with authentication and persistent storage mounted via a specific dedicated local volume. 
All communication between the mobile app and backend is encrypted to ensure that personally identifiable and sensor data are not exposed during transmission. 

The backend architecture is modular and RESTful. It includes user signup and token-based authentication. Each route is organized into its respective controller and router for maintainability and scalability. Data is validated using Mongoose schemas before being saved into MongoDB collections hosted 
on the cloud server. 
Incoming sign-in requests are parsed and validated via express.json() middleware, and routes are protected using JWT-based token authentication. The authMiddleware.js file ensures that only authenticated devices (smartphones) can access or push data to sensitive endpoints. 
For privacy, MotionPI uses JWTs for device-level access control. Each phone is registered with a unique token, stored on-device, and validated on each interaction with the backend. This design for the app from centralized login requirements enables secure offline sign-in. In the token authentication process, the device sends a signup or login request, and then the server verifies the device ID and issues a time-stamped JWT, and then the app stores and attaches this token to all future requests. All requests lacking a valid JWT are rejected with HTTP 401 responses.

Sensitive endpoints perform additional payload validation and format checking to prevent injection attacks or malformed data. This validation includes strict schema-based validation in Mongoose and type checking for arrays, timestamps, and Universally Unique Identifier (UUID). For example, the timestamp is validated to be a floating-point UNIX timestamp, and the phone ID and username fields are required for saving device data. These checks help ensure only valid data is stored.
All data received is saved in MongoDB collections; it stores GPS coordinates, timestamps, phone IDs, device battery percentages, JSON-structured survey responses, and ENMO motion sensor data. Each record includes a timestamp, participant ID, and phone ID, allowing for structured, easy to query storage.

To ensure robustness even in cases of network interruption or app crashes, the app stores a file for unsent data using time-stamped files categorized by data type (e.g., ENMO, surveys), then, using retry mechanisms and scheduled background services, re-sends data when network connectivity is restored. These combined practices ensure both secure and reliable storage and transmission for our system.  Through multiple tests, we have been able to verify that the data that is stored in the 
cloud
database matches the data written directly (and encrypted) to the phone's memory, even when the internet connectivity of the phone had significant and varied interruptions.  

All datasets from the MongoDB cloud database can be directly accessed for batch analysis and visualization via scripts or analytical tools (e.g., Python, R). 
As a result, this system design not only supports longitudinal studies with hundreds of participants, with unique JWTs per phone, but also online and complex analysis.

\section{Ongoing Deployment and Conclusion}
\label{sec:deployment}

\begin{figure}
    \centering
    \includegraphics[width=0.6\linewidth]{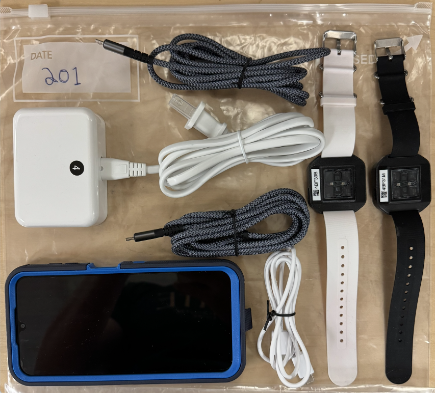}
    \caption{Package of items prepared for each study participant.}
    \label{fig:package}
\end{figure}

The MotionPI infrastructure is currently deployed and in-use as part of an NIH study.  
Each week (up to) 20 
study participants are provided with a smartphone with the MotionPI software installed, along with 2 MotionSense wristbands as shown in Figure \ref{fig:package}.

In ongoing runs, even with continuous BLE connectivity, cellular data, and GPS enabled, the phone and wristbands sustain approximately 3\% battery drain per hour over 14-hour daily windows. This rate is lower than manufacturer endurance estimates for continuous screen-on phone activities such as internet use \cite{samsungA14}. The reduced consumption is attributable to the app’s predominantly screen-off, only event-focused participant interaction, and background data-collection design, which enables a full day of operation. Across participants, the system streams approximately 7.7~million data records per day to the cloud, sufficient for real-time monitoring and later batch analysis.

With minimal instruction, participants wear wristbands, carry the smartphone, and answer the triggered EMA surveys.  The system has been successful in seamlessly and securely gathering all of the requisite data and storing it for easy monitoring and access in the cloud.  This aligns with data more manually recovered from the phone's memory and wristbands.  

We believe that the MotionPI system represents an easy-to-deploy smartphone-centered ecosystem that can facilitate various participatory data collection studies.  It accommodates Bluetooth peripherals, triggered events, EMA surveys, and secure distributed data storage to facilitate simple monitoring and analysis.  Critically, it has important safeguards for data redundancy and privacy protection for personal protected health information.

\bibliographystyle{abbrv}
\bibliography{refs}

\end{document}